\documentclass[aps,pra,twocolumn,amsmath,amssymb,superscriptaddress,floatfix]{revtex4-1}

\usepackage{graphicx}
\usepackage{multirow}
\usepackage{color}
\usepackage{bm}

\def\t#1{\textrm{#1}}

\def\braket#1{\langle #1 \rangle}
\def\n{\nonumber \\ }

\begin{document}

\title{
Photocurrent of exciton polaritons
}

\author{Takahiro~Morimoto}
\affiliation{Department of Applied Physics, The University of Tokyo, Hongo, Tokyo, 113-8656, Japan}
\affiliation{JST, PRESTO, Kawaguchi, Saitama, 332-0012, Japan}

\author{Naoto~Nagaosa}
\affiliation{Department of Applied Physics, The University of Tokyo, Hongo, Tokyo, 113-8656, Japan}
\affiliation{RIKEN Center for Emergent Matter Sciences (CEMS), Wako, Saitama, 351-0198, Japan}

\date{\today}

\begin{abstract}
We study photocurrent of exciton polaritons in inversion broken systems. We use an effective Hamiltonian and Green's function approaches to deduce the formula for polariton photocurrent. The obtained formula shows the polariton photocurrent is proportional to the polariton density and shows opposite signs for the upper and lower branches. Nonvanishing polariton photocurrent requires time reversal symmetry breaking in addition to inversion symmetry breaking. We show that exciton states in transition metal dichalcogenides such as MoS$_2$ support polariton photocurrent once we break time reversal symmetry by applying magnetic fields or using photons with circular polarization. We also perform a simulation based on a rate equation and study a time profile of polariton photocurrent after pulse excitation, which indicates that the photocurrent provides a useful nonlinear probe to study nonequilibrium dynamics of exciton polaritons. 
\end{abstract}

\maketitle

\section{Introduction}

Quantum materials exhibit various interesting nonlinear responses \cite{Boyd,Bloembergen,Sturman}. In particular, the second order nonlinear responses include photocurrent and second harmonic generation which are important both for fundamental physics and technological applications. Photocurrent generation is a recently actively studied topic \cite{Nie,Shi,deQuilettes,Osterhoudt19,Sotome19,Hatada20,deJuan17,Rees20,Nagaosa-review20}. Among various mechanisms for photocurrent generation, shift current is attracting a growing interest~\cite{Kraut,Sipe,Young-Rappe,Morimoto-Nagaosa16,Nagaosa-Morimoto17,Cook17}. Shift current has a geometric origin closely related to the modern theory of polarization. Specifically, shift current arises from the shift of wave function during optical transition between valence and conduction bands. This shift of wave packet is characterized by so called shift vector which is described by Berry connection. While studies on shift current have mainly focused on systems of noninteracting electrons so far, shift current in the presence of electron interactions is an interesting venue, since (collective) excitations in correlated materials generally have large oscillator strength and can potentially enhance the nonlinear functionality. One such example is shift current of excitons which is a bound state of conduction electron and valence hole \cite{Morimoto-exciton16,Chan19}. When an exciton has nonzero polarization due to the shift of wave packet of an electron and a hole, optical excitation of excitons induces an increase of polarization in time, which results in the shift current response. Another example is shift current of magnons in multiferroic materials, where magnon excitation accompanies electric polarization due to the multiferroic nature~\cite{Morimoto-magnon19}. 

\begin{figure}
\begin{center}
\includegraphics[width=0.9\linewidth]{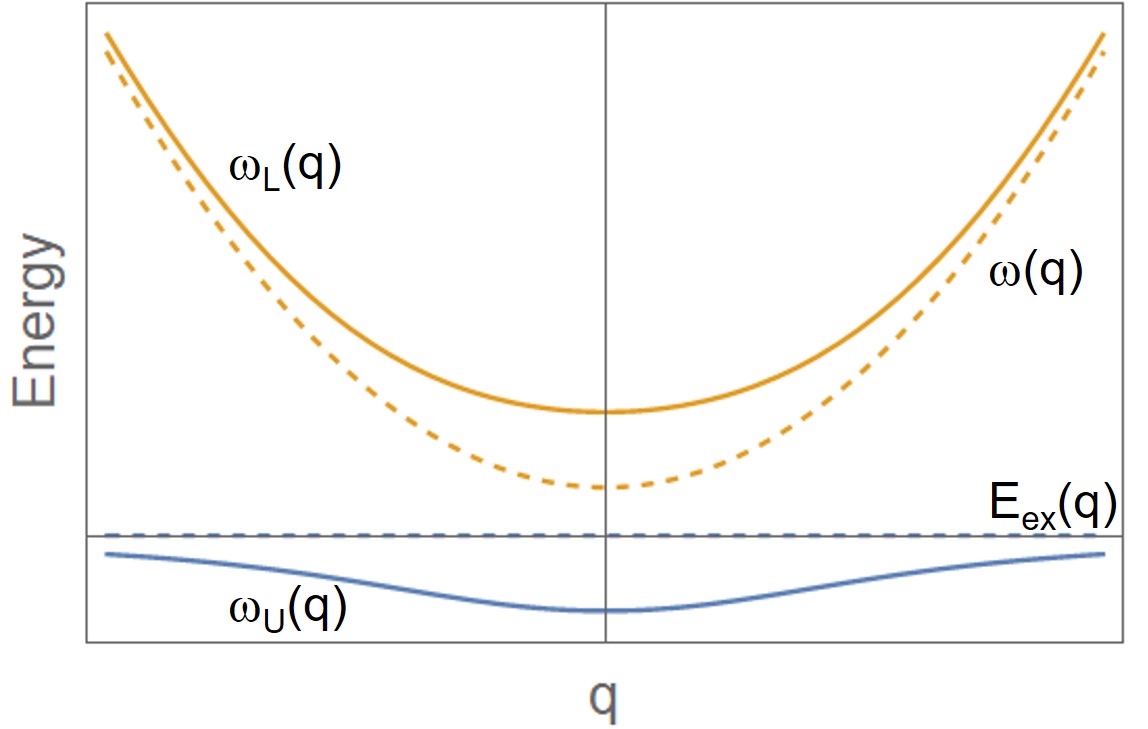}
\caption{\label{fig: polariton band}
A schematic band dispersion of exciton polaritons. 
The exciton state at $E_{ex}(q)$ and the photon at $\omega(q)$ (dashed blue and orange curves) is hybridized to form polariton states with two branches (solid curves). The energy dispersions of the upper and lower polaritons are given by $\omega_U(q)$ and $\omega_L(q)$.  
}
\end{center}
\end{figure}

Exciton polariton is a mixed state of exciton and photon, which generally appears in semiconductor in the presence of light irradiation under strong light-matter coupling.
When the energy levels of photons and the excitons are close with each other, photon and exciton are hybridized and polariton states appear. Namely, polaritons appear due to the anticrossing of a nearly flat band of excitons and a more dispersive band of photons. As a consequence, polariton bands have two branches, called upper branch and lower branch polaritons (Fig.~\ref{fig: polariton band}).
Since polaritons are bosons, they can undergo Bose-Einstein condensation (BEC). BEC of exciton polariton has been experimentally achieved and is an interesting example of macroscopic phase coherence in nonequilibrium composite matter~\cite{YamamotoRMP,Imamoglu96,Kasprzak06,richard2005experimental,lagoudakis2008quantized,kim2011dynamical,Byrnes14}. In the setup for polariton condensates, upper branch polaritons are pumped by light irradiation, which transition into lower branch eventually. Lower branch polaritons are relaxed into the band minimum (at $q=0$) and forms polariton condensate. 
While distribution of polaritons is usually studied from photoemission, it is interesting to seek a possibility of other characterization probes of nonequilibrium dynamics of exciton polaritons. In particular, nonlinear responses of polaritons have not been fully explored so far.

Motivated by these, we study nonlinear responses of exciton polaritons. We focus on photocurrent generation of polariton states. We use an effective Hamiltonian approach and Green's function approach~\cite{Parker19} to study photocurrent of polaritons in the nonequilibrium. We find that nonzero photocurrent appears in inversion broken semiconductors coupled to a cavity. The photocurrent appears from diamagnetic coupling between excitons and cavity photons. The photocurrent turns out to be proportional to the density of polaritons, and shows the opposite sign for upper and lower branch polaritons. Nonvanishing photocurrent requires breaking of time reversal symmetry in addition to broken inversion symmetry. We perform a rate equation analysis for the time profile of polariton density after a pulse excitation, and show that measuring polariton photocurrent gives a useful information about a time profile of the polariton density which reflects detuning between exciton and photon bands and polariton's relaxation paths. Thus, photocurrent of polaritons can be utilized to characterize nonequilibrium dynamics of polaritons.

This paper is organized as follows. In Sec. II, we derive polariton photocurrent based on an effective action. In Sec. III, we present a more detailed analysis of polariton photocurrent using a diagrammatic approach. In Sec. IV, we use a rate equation and study a time profile of polariton photocurrent. In Sec V, we give a brief discussion.

\section{Effective Hamiltonian}
In this section, we derive photocurrent of exciton polariton from an effective Hamiltonian approach.
Let us consider excitons in a cavity described by the Hamiltonian,
\begin{align}
H_0&=\sum_{q} E_{ex}(q) b^\dagger_{q} b_{q} 
+ \sum_{q} \hbar \omega_{q} a^\dagger_{q} a_{q},
\end{align}
where $b_{q}$ and $E_{ex}(q)$ are an annihilation operator and energy dispersion for excitons, and $a_{q}$ and $\hbar \omega_{q}$ are an annihilation operator and energy dispersion for photons, respectively.
We introduce an  effective coupling between the excitons and the electromagnetic field as
\begin{align}
H_{int} &= 
 \sum_{q} [g_1 a^\dagger_q b_q + 
g_2 A a^\dagger_q b_q] + h.c..
\label{eq: ex-ph}
\end{align}
Here the first term is the paramagnetic coupling that mixes the excitons and photons within the rotating wave approximation~\cite{YamamotoRMP}.
The second term is the diamagnetic coupling which is second order in the electromagnetic field. Here we focus on the diamagnetic contribution proportional to the dc component of the electromagnetic field $A$ and the photon field ($a/a^\dagger$) as we study dc current response of exciton polaritons in the following.
Namely, the dc current of the system is given by taking a functional derivative of the Hamiltonian with respect to the (dc) vector potential $A$ as
\begin{align}
\hat J &= \frac{\delta H_{int}}{\delta A}
=\sum_q g_2 a^\dagger_q b_q +h.c. .
\end{align}

The exciton-polaritons appear due to the mixing of the excitons and the photons caused by the $g_1$ term in Eq.~\eqref{eq: ex-ph}. 
We can obtain the polariton operators by diagonalizing $H_0$ and the $g_1$ term as
\begin{align}
a_q &= \alpha b_{U,q} + \beta b_{L,q}, \\
b_q &= - \beta^* b_{U,q} + \alpha^* b_{L,q},
\end{align} 
where $b_U$ and $b_L$ denote annihilation operators of the upper and lower branches of polaritons, respectively, and $\alpha, \beta$ are coefficients satisfying $|\alpha|^2+|\beta|^2=1$.
(We note that the $g_2$ term does not contribute to the polariton formation since the dc component $A$ is infinitesimal.)
Representing $J$ in the polariton basis, we find that the polaritons can induce photocurrent as
\begin{align}
\braket{\hat J} &\simeq
\sum_q 2 \t{Re}[g_2 \alpha^* \beta^*] (\braket{b_{L,q}^\dagger b_{L,q}} - \braket{b_{U,q}^\dagger b_{U,q}}).
\end{align}
Here we neglected cross terms including $b_{U,q}^\dagger b_{L,q}$ and $b_{L,q}^\dagger b_{U,q}$ and only kept terms proportional to the polariton density. This is justified when one can neglect phase coherence between upper branch polariton and lower branch polariton, or when one considers a wave function that consists of a single slater determinant of polariton states.
We can further simplify the expression by using $\alpha^* \beta^* \propto g_1^*/(\hbar \omega_U(q)- \hbar \omega_L(q))$ as
\begin{align}
\braket{\hat J} &\propto \sum_q \frac{\t{Re}[g_1^* g_2]}{\hbar \omega_U(q)- \hbar \omega_L(q)} (\braket{b_{L,q}^\dagger b_{L,q}} - \braket{b_{U,q}^\dagger b_{U,q}}),
\end{align} 
with energies of polaritons in the upper branch $\hbar \omega_U(q)$ and the lower branch $\hbar \omega_L(q)$. 

The above expression clearly shows that nonzero photocurrent appears when polariton states are created and it is proportional to the difference of upper and lower polariton densities.
While the above derivation of the polariton photocurrent is concise and has a clear interpretation, it is based on a rather phenomenological electromagnetic coupling $H_{int}$.
In the next section, we derive the photocurrent of polaritons from a more microscopic Hamiltonian that describes electrons and holes forming the excitons. In particular, we show that the $g_2$ term naturally arises from the diamagnetic coupling of electric fields to the electrons.

\section{Diagrammatic approach}
In this section, we construct a microscopic model of exciton polariton starting from a Hamiltonian that describes interacting electrons and holes, and derive photocurrent of polariton condensate based on a Feynman diagrammatic approach. 

\subsection{Model}
We consider a two band Hamiltonian given by
\begin{align}
H &= H_0 + H_{int} \\
H_0 &= \sum_k (\epsilon_c(k) c_{c,k}^\dagger c_{c,k} 
+\epsilon_v(k) c_{v,k}^\dagger c_{v,k}) \\
H_{int} &= -\sum_{k,k'} V_{k,k'} c_{c,k}^\dagger c_{v,k} c_{v,k'}^\dagger c_{c,k'}
\end{align}
where the subscripts c/v denote conduction and valence bands, and $V_{k,k'}$ is attractive interaction between electrons in the conduction band and holes in the valence band. Hereafter, we set $\hbar=1$ and $e=1$ for simplicity.
We denote the propagators of conduction and valence electrons as
\begin{align}
G_{c/v}(i \omega,k) &= \frac{1}{i\omega - \epsilon_{c/v}(k)}.
\end{align}
To simplify the treatment of exciton formation, we assume a separable form for the attractive interaction as
\begin{align}
V_{k,k'} = w^*(k) w(k'),
\end{align}
where $w(k)$ is some function on the single variable $k$.

We consider a cavity mode of electromagnetic field that has photon dispersion relationship
\begin{align}
H_{ph}&= \sum \omega_{ph}(q) a_q^\dagger a_q
\end{align}
Coupling of electrons to an electric field is given by
\begin{align}
H_{em}= \sum_k A(t) v_{cv}(k) c_{c,k}^\dagger c_{v,k} + h.c.,
\end{align}
with $v_{cv}$ being the interband matrix element of the velocity operator and 
\begin{align}
A(t)&= a_q e^{-i \omega_{ph} t} + a_q^\dagger e^{i \omega_{ph} t}.
\end{align}

\begin{figure}
\begin{center}
\includegraphics[width=\linewidth]{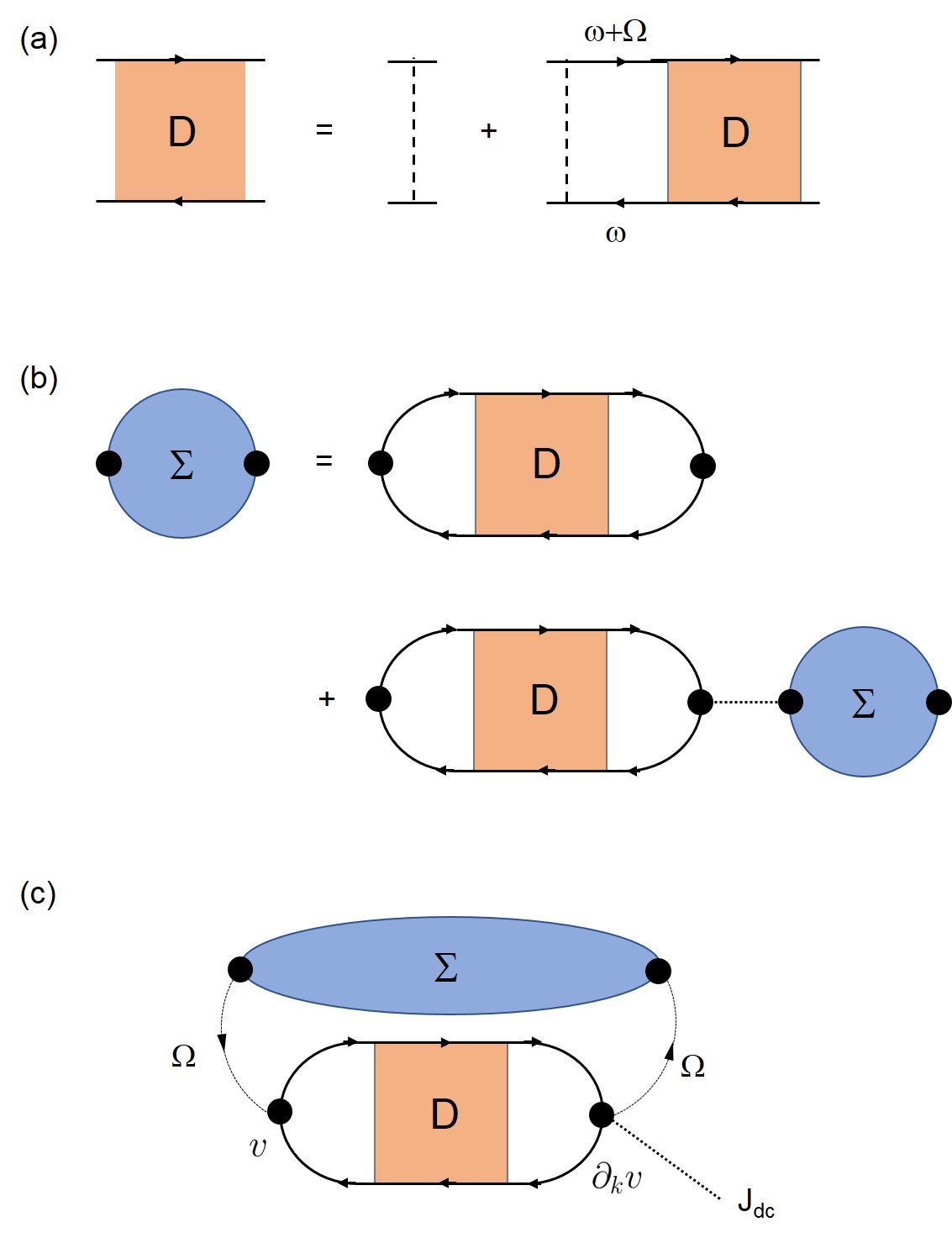}
\caption{\label{fig: diagrams}
Diagrammatic representations for (a) exciton effective interaction $D$, (b) photon self energy $\Sigma$, and (c) polariton photocurrent $J_{dc}$.
}
\end{center}
\end{figure}

\subsection{Exciton propagator}
The exciton propagator consists of a ladder of electron and hole propagators connected with the attractive interaction.
We consider the effective interaction $D_{k,k'}$ where an incoming electron hole pair with the momenta $k'$ is scattered into an outgoing pair with $k$, which is given by
\begin{align}
D_{k,k'}(i\Omega) =& -V_{k,k'} 
 - \int \frac{d\omega}{2\pi} \frac{dk''}{2\pi} D_{k,k''}(i\Omega)G_c(i\omega+i\Omega, k'') \n
&\hspace{5em} \times G_v(i\omega, k'') V_{k'',k'}. 
\end{align}
Taking advantage of the separable form of the interaction, we can write $D_{k,k'}= w^*(k) \tilde D w(k')$ where $\tilde D$ is given by
\begin{align}
\tilde D (i\Omega) &= -1 - \tilde D(i\Omega) \int \frac{d\omega}{2\pi} \frac{dk''}{2\pi} w(k'') G_c(i\omega+i\Omega, k'') \n
&\hspace{7em} \times G_v(i\omega, k'') w^*(k'') \n
&=-1 - \tilde D(i\Omega) \int \frac{dk''}{2\pi}
\frac{|w(k'')|^2}{i\Omega-E_{cv}(k'')}.
\end{align}
with 
\begin{align}
E_{cv}(k)=\epsilon_c(k)-\epsilon_v(k).
\end{align}
This equation can be readily solved as
\begin{align}
\tilde D (i\Omega) &= 
\frac{-1}{1 + \int \frac{dk}{2\pi}
\frac{|w(k)|^2}{i\Omega-E_{cv}(k)}}.
\end{align}
This propagator has a pole below the band gap $\epsilon_c(k)-\epsilon_v(k)$ that corresponds to exciton states. For example, if we take Rice-Mele model $H=t \cos k \sigma_x + \delta t \sin k \sigma_y + m \sigma_z$ and assume $t,\delta t \ll m$, we can approximately write
\begin{align}
\tilde D (i\Omega) &= 
\frac{-1}{1 + \int \frac{dk}{2\pi}
\frac{|w(k)|^2}{i\Omega-2m}} 
= \frac{2m - i\Omega}{i\Omega - (2m - V_{ex})} \n
&\simeq \frac{V_{ex}}{i\Omega - E_{ex}},
\end{align}
where the energy of exciton $E_{ex}$ is given by $E_{ex}=2m-\int \frac{dk}{2\pi} |w(k)|^2$, and $V_{ex}= \int \frac{dk}{2\pi} |w(k)|^2$ corresponds to the exciton binding energy in this case. In the last line, we wrote the singular part at the pole describing the exciton resonance $i\Omega = E_{ex}$.
In general, $\tilde D(i\Omega)$ has a pole structure at exciton excitation below the band gap, and we focus on the corresponding singular part of $\tilde D$ as
\begin{align}
\tilde D(i\Omega) &\simeq \frac{V_{ex}}{i\Omega - E_{ex}},
\end{align}
where $E_{ex}$ is the energy of the exciton and $V_{ex}$ is a constant of the order of exciton binding energy.
Correspondingly, the effective interaction has a pole structure at exciton excitation as
\begin{align}
D_{k,k'}(i\omega) & \simeq \frac{w^*(k)V_{ex} w(k')}{i\omega- E_{ex}}.
\end{align} 

\subsection{Polariton propagator}
We consider polariton which is a mixed state of a photon and an exciton. We can describe polariton excitations by photon self energy.
The photon self energy is given by
\begin{align}
\Sigma(i\Omega) &= \int dk dk' \left[\int d\omega G_v(i\omega, k) v_{vc}(k) G_c(i\omega+i\Omega, k)\right] \n
&\times D_{k,k'}(i\Omega) \left[\int d\omega G_c(i\omega+i\Omega, k') v_{cv}(k') G_v(i\omega, k')\right] \n
&= \int dk dk' \frac{v_{vc}(k)}{i\Omega - E_{cv}(k)} \frac{w^*(k)V_{ex} w(k')}{i\Omega- E_{ex}} \frac{v_{cv}(k')}{i\Omega - E_{cv}(k')},
\end{align} 
as illustrated in Fig.~\ref{fig: diagrams}(b).
If we focus on the exciton resonance at $i\Omega = E_{ex}$,
this can be rewritten in terms of exciton-photon coupling $g_1$ as
\begin{align}
\Sigma(i\Omega) &\simeq \frac{|g_1|^2}{i\Omega- E_{ex}}, \\
g_1 &= \int dk  \frac{\sqrt{V_{ex}} w(k)v_{cv}(k)}{E_{ex}-E_{cv}(k)}.
\label{eq: g1}
\end{align}

We consider photons in cavity that has dispersion relation $\omega_{ph}(q)$. Neglecting $q$ dependence in the exciton dispersion and electron-photon coupling since the scale of the momentum $q$ of photon is much smaller than that of electrons, we obtain the propagator of polariton modes as
\begin{align}
G_p (i\omega, q) &= \frac{1}{i\omega - \omega_{ph}(q) - \Sigma(i\omega)} \n
&= \frac{(i\omega - E_{ex})}{(i\omega - \omega_{ph}(q)) (i\omega - E_{ex}) - |g_1|^2}.
\end{align} 
This propagator has two poles at $i\omega=\omega_U,\omega_L$ with
\begin{align}
\omega_U(q) &= \frac{\omega_{ph}(q) + E_{ex}}{2} + \sqrt{\left(\frac{\omega_{ph}(q) - E_{ex}}{2}\right)^2 + |g_1|^2}, \\
\omega_L(q) &= \frac{\omega_{ph}(q) + E_{ex}}{2} - \sqrt{\left(\frac{\omega_{ph}(q) - E_{ex}}{2}\right)^2 + |g_1|^2}, 
\end{align}
where U/L denotes upper/lower branch of polariton separated by so called LT splitting. 
Focusing on the singular part of the propagator, we can write
\begin{align}
G_p (i\omega, q) &\simeq \frac{\omega_U(q)- E_{ex}}{(\omega_U(q)-\omega_L(q))(i\omega - \omega_U(q))} \n
&+ \frac{E_{ex}-\omega_L(q) }{(\omega_U(q)-\omega_L(q))(i\omega - \omega_L(q))}.
\label{eq: Gp}
\end{align}

\subsection{Photocurrent}
Now we consider photocurrent of polaritons. Polaritons induce dc current via diamagnetic current response as shown in Fig.~\ref{fig: diagrams}(c).
The diamagnetic current is given by
\begin{align}
J &= \sum_k A(t) (\partial_k v)_{cv}(k) c_{c,k}^\dagger c_{v,k} + h.c.
\end{align}
The frequency of current is given by the difference of the energies of particle hole pairs and photons. Polariton excitations are mixed states of particle-hole pair (exciton) and photons where these two energies coincide, and therefore can induce dc photocurrent.

The dc photocurrent is given by 
\begin{align}
J_{dc} &= \int d\Omega dq G_p(i\Omega,q) \n
&
\times \int dk dk' \left[\int d\omega G_v(i\omega, k) v_{vc}(k) G_c(i\omega+i\Omega, k)\right] \n
&\times D_{k,k'}(i\Omega) \left[\int d\omega G_c(i\omega+i\Omega, k') (\partial_{k'} v)_{cv}(k') G_v(i\omega, k')\right] \n
&\simeq  \int d\Omega dq G_p(i\Omega,q)
\left[\int dk \frac{v_{vc}(k)w^*(k)}{E_{ex}-E_{cv}(k)}\right] \frac{V_{ex}}{i\Omega- E_{ex}} \n
&\times \left[\int dk \frac{w(k) (\partial_k v)_{cv}(k)}{E_{ex}-E_{cv}(k)}\right],
\end{align}
where we focused on the exciton resonance at $i\Omega=E_{ex}$ in the last line.
If we define the diamagnetic coupling between exciton and photon as
\begin{align}
g_2 &= \int dk  \frac{\sqrt{V_{ex}} w(k) (\partial_k v)_{cv}(k)}{E_{ex}-E_{cv}(k)},
\label{eq: g2}
\end{align}
we can write the photocurrent as
\begin{align}
J_{dc}&=\int d\Omega dq G_p(i\Omega,q) \frac{g_1^* g_2}{i\Omega - E_{ex}}.
\end{align}
We note that we are interested in polariton excitations at $i\Omega = \omega_{U/L}$, but we used residue at the exciton pole $i\Omega = E_{ex}$ when we simplified the expression using $g_1$ and $g_2$.
This treatment is justified when the exciton binding energy $V_{ex}$ is much larger than the LT splitting of polaritons.
Polariton photocurrent is obtained by keeping poles describing two branches of polaritons in Eq.~\eqref{eq: Gp}, which yields
\begin{align}
J_{dc} 
&\simeq \int d\Omega dq \frac{g_1^* g_2}{(\omega_U(q)-\omega_L(q))}
\left[\frac{1}{i\Omega - \omega_U(q)} - \frac{1}{i\Omega - \omega_L(q)} \right].
\end{align}
For finite temperatures, we replace the $\Omega$ integral with Matsubara frequency summation, where the poles give occupation numbers of corresponding modes. Thus the dc current is given by
\begin{align}
J_{dc} &= \int dq \frac{g_1^* g_2}{(\omega_U(q)-\omega_L(q))} (n_U(q) - n_L(q)), 
\label{eq: polariton photocurrent}
\end{align}
with occupation numbers (Bose distribution functions) $n_U(q)$ and $n_L(q)$ for upper and lower polariton branches with momentum $q$.

The photocurrent of polaritons in the nonequilibrium systems are also given by Eq. \eqref{eq: polariton photocurrent}. 
In polariton condensates, the system is at the nonequilibrium state under pumping with external laser light. In this case, we can use Keldysh Green's function instead of Matsubara Green's functions and the current response is obtained by taking the lesser component of the diagram~\cite{Rammer86,Jauho94}. Since we are interested in polariton excitations, we just need to replace polariton propagator $1/(i\Omega-\omega_{U/L}(q))$ with its lesser component. After $\Omega$ integration with using Keldysh equation, this procedure just gives the occupation numbers of polaritons $n_U$ and $n_L$ in the nonequilibrium states.
Therefore Eq. \eqref{eq: polariton photocurrent} still holds in the nonequilibrium situations including polariton condensates. 

Equation \eqref{eq: polariton photocurrent} shows that the photocurrent is opposite for upper and lower polariton branches and the current response is enhanced at momentum $q$ where the splitting  $\omega_U(q)-\omega_L(q)$ is small, which is a useful feature in deducing nonequilibrium dynamics of polaritons from a time profile of the photocurrent, as we discuss in the next section.

\subsection{Symmetry consideration \label{sec: symmetry}}
We present symmetry consideration of the polariton photocurrent. We consider effects of inversion symmetry $\mathcal{I}$ and time reversal symmetry $\mathcal{T}$. According to Eq.~\eqref{eq: polariton photocurrent}, the polariton photocurrent requires that two coupling constants $g_1$ and $g_2$ are both nonzero. 
Based on the expressions Eq.~\eqref{eq: g1} and Eq.~\eqref{eq: g2} for $g_1$ and $g_2$, the symmetry properties of these coupling constants can be deduced from the relationships under $\mathcal{I}$:
\begin{align}
 E_{cv}(k) &\to E_{cv}(-k) \\
 v_{cv}(k) &\to -v_{cv}(-k) \\
 \partial_k v_{cv}(k) &\to \partial_kv_{cv}(-k) \\
 w(k) &\to \eta_I w(-k)
\end{align}
and the relationship under $\mathcal{T}$:
\begin{align}
 E_{cv}(k) &\to E_{cv}(-k) \\
 v_{cv}(k) &\to -v_{cv}^*(-k) \\
 \partial_k v_{cv}(k) &\to \partial_kv_{cv}^*(-k) \\
 w(k) &\to \eta_T w^*(-k)
\end{align}
with $\eta_I=\pm 1$ and $\eta_T=\pm 1$.
These relationships can be obtained from the symmetry actions on electron operators
\begin{align}
\mathcal{I}:\qquad & c_{c/v,k} \to  c_{c/v,-k},
\end{align}
and
\begin{align}
\mathcal{T}:\qquad & c_{c/v,k} \to  c_{c/v,-k},
\end{align}
with complex conjugation on c-numbers ($\mathcal{T}=\mathcal{K}$).
We note that the signs $\eta_I$, $\eta_T$ arise from the fact that the combination $w^*(k)w(k')$ should be invariant under $\mathcal{I}$ and $\mathcal{T}$.

First we consider the effect of inversion symmetry $\mathcal{I}$. 
When $\eta_I=+1$, the integrand in Eq.~\eqref{eq: g1} is odd with respect to $k$, and we obtain $g_1=0$ after integration. Vanishing $g_1$ means that excitons and photons do not couple with each other and polaritons are not formed, which is a situation that we are not interested in.
When $\eta_I=-1$, the integrand in Eq.~\eqref{eq: g1} is even with respect to $k$, and we obtain nonzero $g_1$ after integration, where polaritons are formed as expected. In this case, Eq.~\eqref{eq: g2} shows that its integrand becomes odd and $g_2$ vanishes. Therefore, we find polariton photocurrent vanishes in the presence of inversion symmetry.
This is natural in that the direction of photocurrent cannot be determined when the system is $\mathcal{I}$ symmetric.

Next we consider time reversal symmetry $\mathcal{T}$.
When $\eta_{T}=+1$, $g_1$ is pure imaginary and $g_2$ is real.
When $\eta_{T}=-1$, $g_1$ is real and $g_2$ is pure imaginary.
Thus $\t{Re}[g_1^* g_2]=0$ in both cases which results in vanishing polariton photocurrent.

Therefore measuring polariton photocurrent requires that both inversion and time reversal symmetries are broken. Inversion breaking can be achieved by using noncentrosymmetric crystals or designing the quantum well for excitons in a noncentrosymmetric way. 
Broken $\mathcal{T}$ symmetry can be implemented, for example, by applying magnetic field or introducing magnetic impurity to induce spin splitting. Another possibility to break $\mathcal{T}$ is to use cavity photons in a circularly polarized state.

\begin{figure}[tb]
\begin{center}
\includegraphics[width=0.9\linewidth]{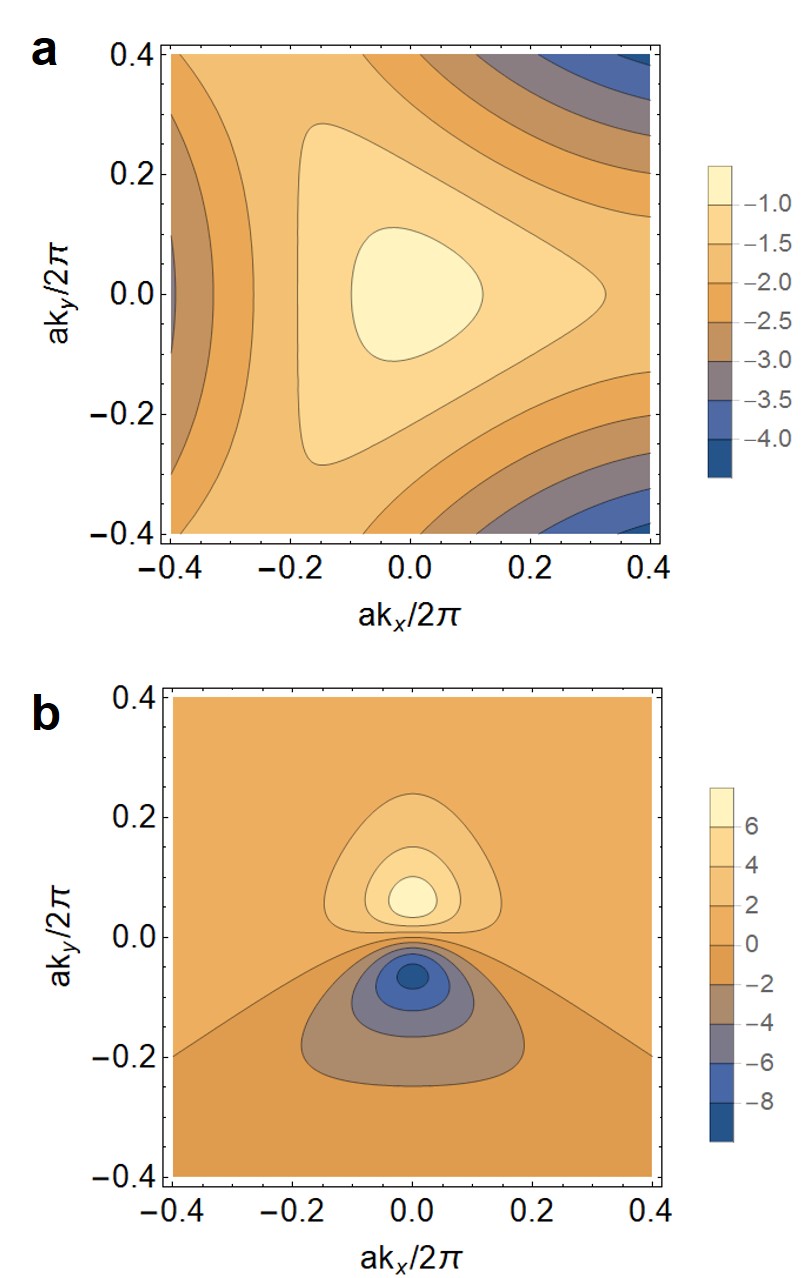}
\caption{\label{fig: tmd}
(a) Contour plot of energy dispersion of the valence band around the K point obtained from the two band model for MoS$_{2}$. The trigonal warping is introduced through the $g_3$ term.
(b) Contour plot of integrand for $g_{1,x}$ in Eq.~\eqref{eq: g1} around the K point, where $k_x$ and $k_y$ are measured from K point.
}
\end{center}
\end{figure}

\subsection{Application to transition metal dichalcogenides}
In this subsection, we apply our theory of polariton photocurrent to transition metal dichalcogenides (TMDs), especially, focusing on the case of MoS$_2$.
TMDs exhibit large exciton binding energy and would provide a suitable platform for exciton polaritons and their photocurrent. 
We adopt an effective two band model for TMDs around K/K' points, given by
\begin{align}
H=
\begin{pmatrix}
\Delta/2 + \gamma_1^2 a^2k^2  & t ak_- + \gamma_3 a^2k_+^2 \\
t ak_+ + \gamma_3 a^2k_-^2 &  -\Delta/2  + \tau_z s_z \lambda + \gamma_2^2 a^2k^2 \\
\end{pmatrix}.
\end{align}
Here, $k_\pm = \tau_z k_x \pm i k_y$ with $\tau_z=\pm 1$ for K and K' valleys, and $k^2=k_x^2+k_y^2$. $\Delta-\lambda$ is the band gap with spin orbit coupling $\lambda$ and $s_z=\pm 1$ for spins, where we consider the bands with $s_z=\tau_z$ since we are interested in electronic structure near the band gap. $\gamma_1, \gamma_2$ and $\gamma_3$ are coefficients of $k^2$ terms. In particular, $\gamma_3$ introduces trigonal warping of the band structure. We use parameters for MoS$_2$ obtained from band fitting to first principle calculations in Ref.~\cite{Liu13}: 
$a=3.190 \t{\AA}, \Delta=1.663\t{eV}, t=1.059\t{eV}, \gamma_1=0.055\t{eV}, \gamma_2=0.077\t{eV}, \gamma_3=-0.123\t{eV}, \lambda=0.073\t{eV}$.
The band structure of the valence band is shown in Fig.~\ref{fig: tmd}(a).
For the exciton binding energy, we adopt the experimentally reported value $\Delta-\lambda-E_{ex}=0.44\t{eV}$ \cite{Hill15}.
Since the above two band model is only valid around K/K' points, we set momentum cutoff $\Lambda=0.8\pi/a$ and perform $k$ integration in the region $-\Lambda<k_x\le \Lambda$ and $-\Lambda<k_y\le \Lambda$.
We consider the interaction between the electrons and holes given in the form
\begin{align}
w(k) =\sqrt{U} 
\begin{pmatrix}
1 & 0 \\
0 & -1 \\
\end{pmatrix},
\end{align}
which comes from the Hartree term of the onsite interaction $U$ which we set $U=3$eV. (While we also have the term proportional to the identity, it does not contribute to the interaction between conduction and valence electrons due to the orthogonality of wave functions.) 

As discussed in Sec.~\ref{sec: symmetry}, nonzero polariton photocurrent requires time reversal symmetry breaking in addition to the inversion symmetry breaking from the noncentrosymmetric crystal structure of TMDs. We consider two possibility of $\mathcal{T}$ breaking: (i) application of magnetic fields/proximity to magnetic materials, (ii) usage of cavity photons with circular polarization.

In the case (i), we consider application of magnetic fields or proximity to magnetic materials which introduce asymmetry between K and K' valleys. Namely, application of large magnetic field is reported to modulate the band gaps at K and K' valleys differently~\cite{Stier16}.
In such cases, we can focus on exciton states at only one valley and neglect contributions from the other valley, due to the energy difference of the exciton states.
Since TMDs are 2D materials, we consider the 2D version of the coefficients in Eqs.~\eqref{eq: g1} and \eqref{eq: g2} as
\begin{align}
g_{1,\alpha} &= \int dk  \frac{\sqrt{V_{ex}} w_{vc}(k)(v_\alpha)_{cv}(k)}{E_{ex}-E_{cv}(k)}, \\
g_{2,\beta\gamma} &= \int dk  \frac{\sqrt{V_{ex}} w_{vc}(k) (\partial_{k_\beta} v_\gamma)_{cv}(k)}{E_{ex}-E_{cv}(k)},
\end{align}
with the velocity operator $v_\alpha$ along the $\alpha$ direction, where the subscripts $\alpha, \gamma$ specify the polarization of photons and $\beta$ specifies the direction of the photocurrent.
By applying Eq.~\eqref{eq: g1} and Eq.~\eqref{eq: g2}, we obtain the coefficients as
$g_{1,x}=-2.9\tau_z ~\t{meV}, g_{1,y}=6.8i~\t{meV}$ and 
$g_{2,xx}=16\tau_z ~\t{meV\AA}, g_{2,yy}=13~\t{meV\AA}$.
The integrand for $g_{1,x}$ is shown in Fig.~\ref{fig: tmd}(b).
This result indicates that polariton photocurrent along the $x$ direction $J_x$ with $x$ polarized photons has nonzero but opposite contributions from the two valleys ($\propto\t{Re}(g_{1,x}g_{2,xx})$). Hence, introducing asymmetry between the two valleys and selectively creating the exciton state at one valley results in nonzero polariton photocurrent.
We note that the photocurrent along the $y$ direction with $y$ polarized photons vanishes due to $\t{Re}(g_{1,y}g_{2,yy})=0$, which is a consequence of the unbroken $\mathcal{T}R_x= \mathcal{K}$ satisfying $\mathcal{T}R_x H(k_x, -k_y) (\mathcal{T}R_x)^{-1}=H(k_x, k_y)$.

In the case (ii), the cavity photons are circularly polarized and break $\mathcal{T}$ symmetry. Cavity photons with circular polarization is realized, e.g, in Refs.~\cite{Martin02,Sarkar10}, and would be applicable to exciton polaritons in TMDs.
We consider the coefficients under left/right circularly polarized light: $g_{1,L/R}$ with $v_x \pm i v_y$ in Eq.~\eqref{eq: g1}, and $g_{2,\alpha L/R}$ with $\partial_{k_\alpha}(v_x \pm i v_y)$ in Eq.~\eqref{eq: g2}. 
By applying Eq.~\eqref{eq: g1} and Eq.~\eqref{eq: g2} and summing contributions from the two valleys, we obtain the coefficients as
$g_{1,L}=-14 ~\t{meV}$ and 
$g_{2,xL}=34 ~\t{meV\AA}, g_{2,yL}=26i~\t{meV\AA}$.
In the case of circularly polarized light, contributions from the two valley do not cancel with each other, and we obtain photocurrent along the $x$ direction ($\propto \t{Re}(g_{1,L}g_{2,xL})$), 
while the photocurrent along the $y$ direction vanishes ($\t{Re}(g_{1,L}g_{2,yL})=0$).
This behavior is consistent with the symmetry property of the second order nonlinear coefficient $J_\alpha = \sigma_{\alpha\beta\gamma}E_\beta E_\gamma$, where $R_x$ symmetry indicates $\sigma_{xxy} \neq 0$ and $\sigma_{yxy} = 0$ (Note that the response to the circularly polarized light is given by $J\propto E_xE_y$).  

The above calculations demonstrate that TMDs can support nonzero polariton photocurrent once the $\mathcal{T}$ symmetry is broken. In particular, large exciton binding energy in TMDs and their controllability suggests that TMDs are candidate materials for observing polariton photocurrent.

\begin{figure}[tb]
\begin{center}
\includegraphics[width=0.9\linewidth]{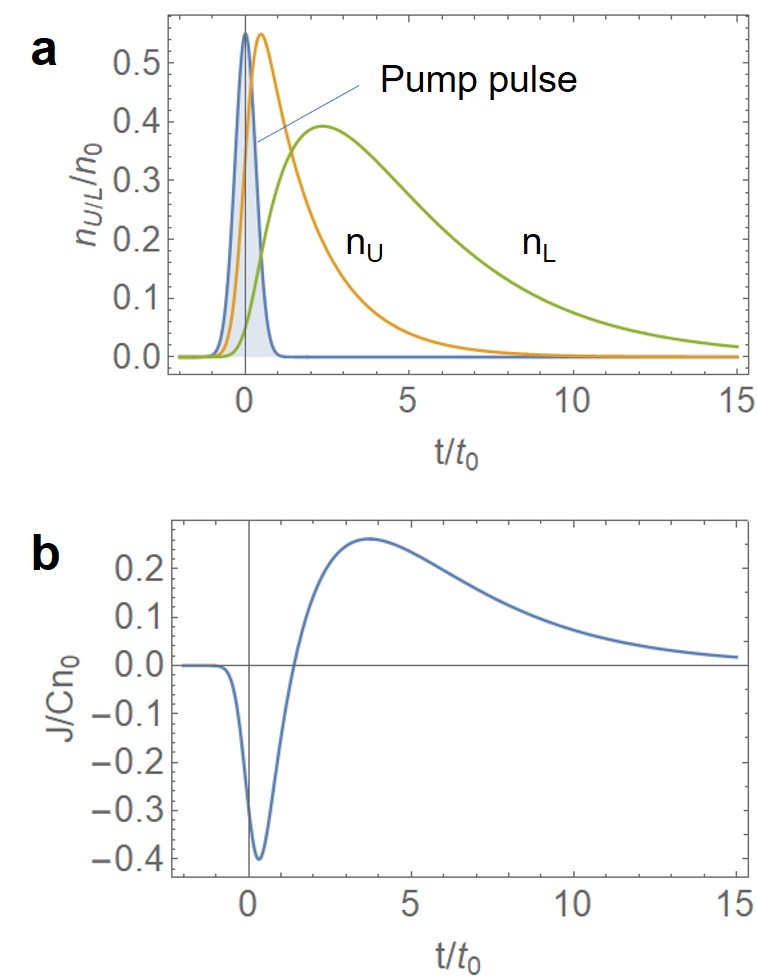}
\caption{\label{fig: polariton density}
A solution of the rate equation for polariton density and photocurrent. 
(a) Time profile of polariton density after pulse pumping.
Polariton is created in the upper branch by the pump pulse.
Polariton density in the upper branch $n_U$ decreases due to the transitioning into the lower branch with the rate $r_1$. 
Correspondingly, the polariton density in the lower branch $n_L$ increases initially and decays eventually due to the decay of the polaritons with the rate $r_2$. 
(b) Time profile of polariton photocurrent. 
Photocurrent grows quickly after pulse pumping due to population of upper branch polaritons. Afterwards, the photocurrent shows a characteristic sign change due to transition of polaritons from the upper branch to the lower branch, reflecting the fact that polariton photocurrent is proportional to the difference of polariton densities of the two branches.
Here we adopted the parameters $p(t)=e^{- 5(t/t_0)^2}n_0/t_0$ and $r_1=0.6/t_0, r_2=0.3/t_0$ with the typical polariton density $n_0$ which ranges from $10^9\t{ cm}^{-2}$ to $10^{11}\t{ cm}^{-2} $, and the characteristic time scale $t_0$ of the order of 1 - 10 ps.
}
\end{center}
\end{figure}

\begin{figure*}
\begin{center}
\includegraphics[width=0.9\linewidth]{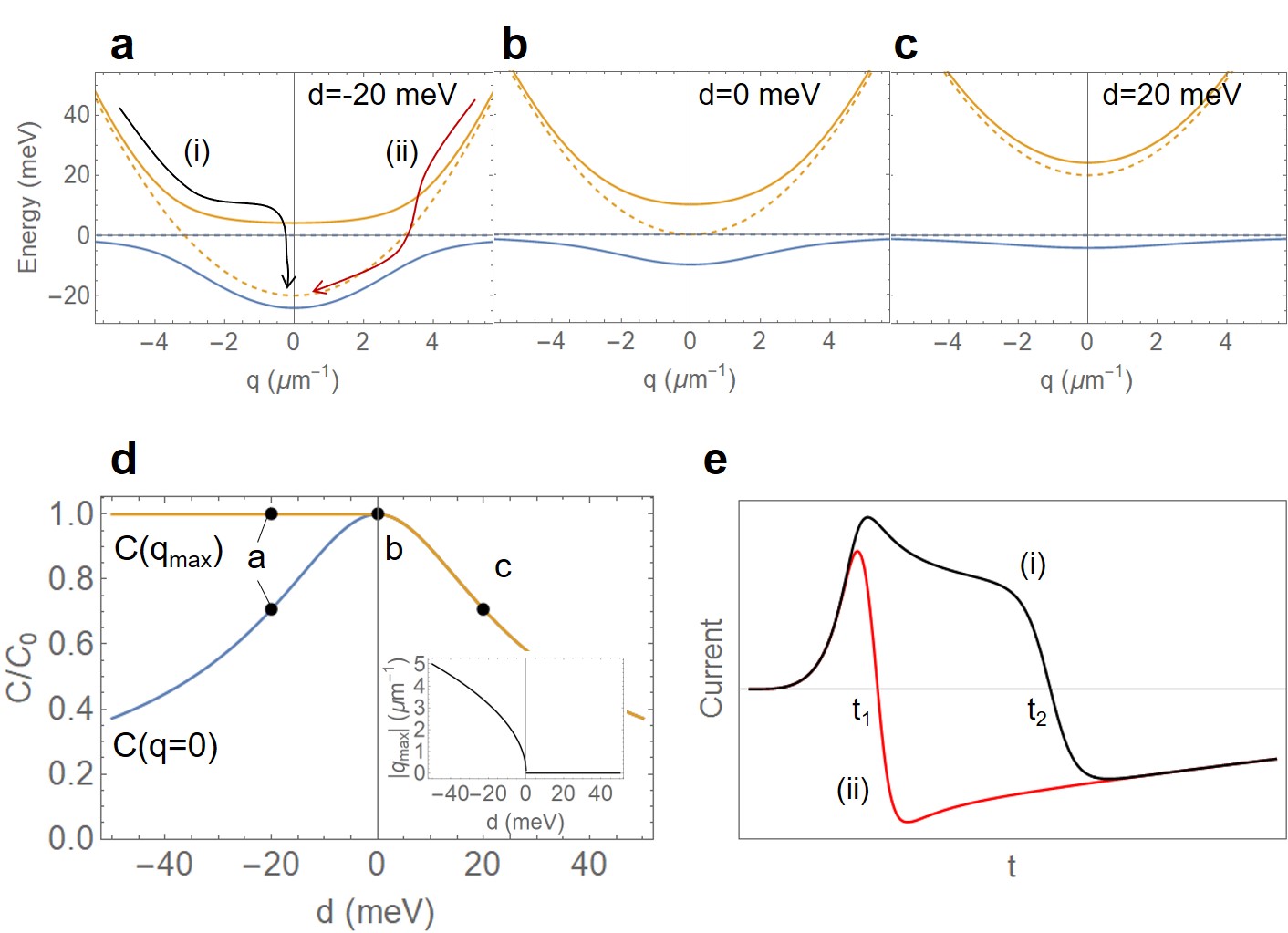}
\caption{\label{fig: d dependence}
Detuning dependence of polariton photocurrent. 
(a,b,c) Band structures of polaritons for several detunings between the exciton state and the photon dispersion.  
We set the exciton energy $E_{ex}=0$ and consider the photon energy dispersion $\hbar\omega(q)=a q^2+d$ with the coupling $g_1$. 
We adopted the parameters $a=2 \t{~}\mu\t{m}^2\t{meV}$, $g_1=10$ meV, and the energy detunings (a) $d=-20$ meV, (b) $d=0$ meV, and (c) $d=20$ meV.
(d) Coefficient $C$ for the polariton photocurrent plotted as a function of detuning $d$. Blue and orange curves represent $C(q=0)$ and $C(q_\t{max})$, respectively, where $q_\t{max}$ is the momentum where band splitting becomes smallest. The inset shows $q_\t{max}$ as a function of $d$. 
The photocurrent coefficient $C$ shows a peak structure at the zero detuning for $q=0$. The coefficient $C(q_\t{max})$ is constant for negative detuning ($d<0$) since the anticrossing of the two bands occurs at $q_\t{max}$.
 The black dots represent the detunings corresponding to the band structures in (a,b,c).
We used the normalization constant $C_0=e g_2/\hbar^2$.
(e) Schematic time profiles of the photocurrent that correspond to the two representative relaxation paths of polaritons, (i) and (ii) shown in (a). 
The photocurrent takes maximum when the polariton is at $q_\t{max}$ (at the time $t_1$), and shows a sign change when polaritons transition into the lower branch (i) at the later time $t_2$ and (ii) right after $t_1$. $t_2$ is the time when the polariton transitions from the bottom of the upper branch to the lower branch in the path (i).
}
\end{center}
\end{figure*}

\section{Rate equation}
In this section, we consider a time profile of photocurrent of the exciton polariton. To do so, we combine the obtained expression in the previous section with the rate equation to account for the time dependence of polariton occupation. 
We consider a simple rate equation that involves pumping of polariton into the upper branch ($p(t)$), transition from the upper branch to the lower branch ($r_1$), and decay from the lower branch ($r_2$).
The rate equation reads
\begin{align}
\frac{d n_U}{dt} &= p(t) - r_1 n_U, \\
\frac{d n_L}{dt} &= r_1 n_U - r_2 n_L,
\end{align}
for polariton density $n_{U/L}$ for the upper/lower branch polaritons.
The photocurrent of polariton is given by
\begin{align}
J(t) &= C (n_L(t)-n_U(t)),
\end{align}
with 
\begin{align}
C=\frac{e g_1^* g_2}{\hbar^2 (\omega_U-\omega_L)}.
\label{eq: C}
\end{align}
Here we ignore effects of band dispersion of the polariton branches, and use the coupling constants $g_1$ and $g_2$ at $q=0$ and the energy splitting $\omega_U(q=0)-\omega_L(q=0)$. Also, we restored $e$ and $\hbar$ in the above expression.
 
Figure \ref{fig: polariton density} shows the time profile of polariton density and photocurrent under pulse pumping at $t=0$.
Here we adopted the parameters $p(t)=e^{- 5(t/t_0)^2}n_0/t_0$ and $r_1=0.6/t_0, r_2=0.3/t_0$ with the typical polariton density $n_0$ which ranges from $10^9\t{ cm}^{-2}$ to $10^{11}\t{ cm}^{-2} $, and the characteristic time scale $t_0$ of the order of 1 - 10 ps.
The pump pulse creates polariton in the upper branch and $n_U$ increases quickly at the initial time. The upper branch polaritons undergo transition into the lower branch polaritons, and $n_L$ increases afterwards and becomes larger than $n_U$ eventually. Finally both $n_U$ and $n_L$ decrease due to the decay of polaritons from the lower branch.
Correspondingly, the photocurrent shows a characteristic time profile. The photocurrent $J(t)$ quickly grows after the pump pulse as the upper branch polaritons are created. 
Since $J(t) \propto n_L(t) - n_U(t)$, the photocurrent shows a characteristic sign change due to the transition of polaritons from the upper branch to the lower branch. At the later time, $J(t)$ diminishes, according to the decay of polaritons from the lower branch.
Such characteristic time profile of photocurrent can be used to deduce the polariton density in each branch by measuring photocurrent.
In particular, when the polariton splitting is smallest at $q=0$, the high density of $n_L(q=0)$ due to the polariton condensation can lead to large photocurrent according to Eq.~\eqref{eq: polariton photocurrent}. This implies that polariton photocurrent can signal the polariton condensation in such polariton dispersions.

Next let us consider how the photocurrent behaves for different types of anticrossing of exciton and photon bands (i.e., for different values of detuning between those two bands), by incorporating $q$ dependence of polariton dispersion.
In order to discuss the time profile of the photocurrent through the relaxation of polaritons after its creation, we suppose that the polariton is located at the momentum $q(t)$ at the time $t$, and we consider the photocurrent
\begin{align}
J(t) &= C(q(t)) (n_L(t)-n_U(t)),
\label{eq: J(t)}
\end{align}
with the $q$ dependent coefficient
\begin{align}
C(q)=\frac{e g_1^* g_2}{\hbar^2 (\omega_U(q)-\omega_L(q))}.
\label{eq: C}
\end{align}
Here we neglect $q$ dependence in the coupling constant $g_1$ and $g_2$, while we restore the $q$ dependence in the energy splitting in the denominator.
Figures \ref{fig: d dependence}(a,b,c) show the band structures of exciton polaritons for different detunings between the exciton state and the photon dispersion. We set the exciton energy $E_{ex}=0$ and consider the photon energy dispersion $\hbar\omega(q)=a q^2+d$ with the detuning $d$.
We adopted the parameters $a=2 \t{~}\mu\t{m}^2\t{meV}$ and $g_1=10$ meV.
Figure \ref{fig: d dependence}(d) shows the coefficient $C$ for the polariton photocurrent as a function of the detuning (with the normalization constant $C_0=e g_2/\hbar^2$). The blue curve represents the coefficient $C(q=0)$ for the polaritons at $q=0$ states, and the orange curve for the polaritons at the momentum $q_\t{max}$ where energy splitting is the smallest. 
Since the coefficient $C$ is inversely proportional to the splitting of two branches, $C(q=0)$ shows a peak structure at zero detuning, and decays as the magnitude of the detuning $|d|$ increases. 
This is reasonable because the polariton photocurrent arises from (diamagnetic) coupling between excitons and photons, and the coupling becomes most efficient when these two satisfy the resonant condition at $d=0$. This feature indicates that the magnitude of photocurrent gives an information about the relative position of exciton and photon bands.      
In contrast, $C(q_\t{max})$ depicted with the orange curve in Fig.~\ref{fig: d dependence}(d) is the maximum of $C(q)$ for given $d$, where $q_\t{max}$ is the momentum given by $\omega(q_\t{max})=E_\t{ex}$ which maximizes $C(q)$.  While $C(q_\t{max})$ coincides with $C(q=0)$ for positive detuning, $C(q_\t{max})$ becomes constant for negative detuning since the energy splitting at the anticrossing at $q_\t{max}$ is constant with $2g_1$. This indicates that the photocurrent becomes largest when the polaritons go through the anticrossing at $q=q_\t{max}$ in the course of relaxation, because of $1/(\omega_U(q)-\omega_L(q))$ factor in Eq.~\eqref{eq: C}.

Finally, let us discuss the time profile of photocurrent in the relaxation process of polaritons after excitation.
For example, we consider two representative relaxation paths depicted as (i) and (ii) in Fig.~\ref{fig: d dependence}(a), where corresponding time profiles of the photocurrent are illustrated schematically in Fig.~\ref{fig: d dependence}(e). 
In the path (i), polaritons relaxes within the upper branch toward the band bottom and transition into the lower branch. In this case, the photocurrent shows peak structure at $t=t_1$ when the polariton is at the anticrossing point ($q=q_\t{max}$), and shows a sign change upon the transition into the lower branch at later time $t_2$ when the polariton reaches the band bottom.
In the path (ii), polaritons transition into the lower branch at the anticrossing point. Accordingly, the photocurrent changes sign at $t=t_1$.
Thus, measuring the polariton photocurrent provides an information about the relaxation path with its magnitude and sign.

\section{Discussions}
We have shown that photocurrent from polaritons appears when inversion symmetry is broken due to diamagnetic coupling between electrons and photons. Since photocurrent generation leads to nonzero voltage at the boundary of the sample, polariton photocurrent generates nonzero power when the sample is connected to electrodes. This appearance of nonzero power requires energy supply. For example, in the case of shift current photovoltaics, the energy supply comes from the absorption of photons that creates electron hole pairs across the band gap.
In the present case, the energy supply comes from photo-creation of exciton polariton states. Polaritons carrying photocurrent are dissociated into electron-hole pairs once they reach the electrodes. In the steady state with finite dc current flowing, one needs to supply polaritons constantly which involves absorption of photons and behaves as a source of energy supply. 


While we mainly considered polaritons realized with cavity photons in the setup for polariton condensate, polaritons generally appear in bulk crystals. Therefore, we can also expect photocurrent generation in bulk noncentrosymmetric crystals when polariton modes are excited.
For example, GaAs and AlAs would be candidate materials for observing polariton photocurrent in the bulk crystals~\cite{schaefer1997nonlinear,tsintzos2009room}.

Finally let us perform a crude estimation of the order of magnitude of polariton photocurrent. 
According to Eq.~\eqref{eq: J(t)}, the photocurrent is given by 
\begin{align}
J &= C n \simeq \frac{e g_1 g_2}{\hbar^2(\omega_U-\omega_L)} n 
\simeq \frac{e g_1  R n}{\hbar}.
\end{align}
Here we used an estimation for the diamagnetic coupling $g_2 \simeq g_1 R$ where $R$ is the so called shift vector of the order of lattice constant ($R \simeq 1 \AA$).
The energy splitting of the two branches is given by $g_1 \simeq 10$ meV and the polariton (2D) density is given by $n \simeq 10^9\t{ cm}^{-2}$ ~\cite{YamamotoRMP}.
These values lead to an estimate for the current density as $J\simeq 3\times 10^{-3} \t{ A/m}$.
Since the size of polariton condensate is of the order of 10 $\mu$m, this amounts to the photocurrent of the order of 30 nA, which is feasible for measurement.

\begin{acknowledgements}
We thank Yoshinori Tokura for fruitful discussions.
This work was supported by The University of Tokyo Excellent Young Researcher Program, JST PRESTO (JPMJPR19L9), JST CREST (JPMJCR19T3)(TM), and
JST CREST (JPMJCR1874 and JPMJCR16F1), 
JSPS KAKENHI (18H03676 and 26103006)(NN).
\end{acknowledgements}

\bibliography{reference}

\end{document}